\documentclass[aps,pra,twocolumn,groupedaddress]{revtex4-1}

\usepackage{mathtools}
\usepackage{amssymb}
\usepackage{pgfplots}
\usepackage[justification=raggedleft]{subcaption}
\usepackage[justification=raggedright]{caption}
\begin{document}


\title{Weak three-dimensional mediators of two-dimensional triplet pairing}

\author{Shane Kelly}
\email[]{shane.kelly@ucr.edu}
\author{S.-W. Tsai}
\affiliation{Department of Physics and Astronomy, University of California, Riverside, CA 92521, USA}

\date{\today}

\begin{abstract}
    Recent experiments demonstrate the ability to construct cold atom mixtures with species selective optical lattices.
    This allows for the possibility of a mixed-dimension system, where one fermionic atomic species is confined to a two dimensional lattice, while another species is confined to a three dimensional lattice that contains the two-dimensional one.
    We show that by tuning the density of an arbitrary number of three-dimensional atomic species, we can engineer an arbitrary, rotationally-symmetric, density-density, effective interaction for the two-dimensional particles.
    This possibility allows for an effective interaction that favours triplet pairing for two-dimensional, $SU(2)$ symmetric particles.
    Using a functional renormalization-group analysis for the two-dimensional particles, we derive and numerically confirm that the critical temperature for triplet pairing depends exponentially on the effective interaction strength.
    We then analyse how the stability of this phase is affected by the particle densities and the fine tuning of interaction parameters.
    We conclude by briefly discussing experimental considerations and the potential to study triplet pairing physics, including Majorana fermions and spin textures, with cold atoms on optical lattices.
\end{abstract}


\maketitle

\section{Introduction}
A remarkable result of quantum mechanics is that the properties of low temperature liquids can be described by a single wave function.
This phenomenon, known as superfluidity, can present itself in both fermionic and bosonic systems.
In the simplest bosonic system, this wave function has only one degree of freedom associated with the superfluid phase.
In fermionic systems where condensation occurs due to pairing, the wave function can have additional degrees of freedom associated with the spin and angular momentum of the pairing state.
In the most common case, the odd exchange symmetry for the fermions is satisfied by the singlet spin state.
In two dimensions, the remaining orbital degree of freedom is then fixed perpendicular to the plane -- the degrees of freedom of the superfluid phase again being reduced to one.
More complex states are possible when the odd exchange symmetry is satisfied by an odd orbital wave function and the wave function is degenerate in the three triplet states.
These states tend to be less common in spin-half systems since they carry higher angular momentum than the isotropic $s$-wave singlet state, and thus tend to be energetically less favorable.

Superfluid $^3$He, a fermionic superfluid with $p$-wave orbital pairing, avoids this tendency by an anisotropic van der Waals interaction.
The spin degree of freedom then gives rise to various interesting spin textures, vortices and other unusual properties \cite{Leggett1975}. 
A particularly interesting property occurs when breaking the degeneracy of the spin triplet state by spin-orbit coupling.
In this case, the vortices that arise as the temperature is increased carry an angular momentum with half a flux quantum.
These vortices are known to host Majorana fermions that carry non-Abelian statistics and have been used in proposals for quantum computing \cite{Tewari2007,Alicea2012}.

These exotic phases, and the possibility of employing them for quantum computing, have lead to a large search for $p$-wave superfluids and Majorana fermions.
One of the earliest examples since $^3$He was in the superconducting phase of $Sr_2RuO_4$ \cite{Maeno2012,Kallin2016}.
Cold atoms have a particular lure, because they are highly controllable, and they offer the possibility of studying condensation in a weak-coupling limit where perturbative results apply.
Proposals for $p$-wave superfluidity in cold atoms range in mechanism, including taking advantage of anisotropic effects due to a dipolar interaction \cite{Baranov2002,Bruun2008,Bhongale2012,Bhongale2013}, using long-range interactions due to a bosonic mediator \cite{Efremov2002,Mathey2006,Mathey2007} and using $p$-wave Feshbach resonances \cite{Gurarie2005,Cheng2005,Zhang2008}.

More recently, species-specific optical lattices \cite{Leblanc2007} have been experimentally implemented \cite{Lamporesi2010a} and have allowed for systems where one species of atom is confined to a plane or a wire, while the other species is free to explore a 3D trap.
This possibility led to a new set of proposals \cite{Nishida2009, Wu2016,Midtgaard2016} for $p$-wave pairing mechanisms in cold atom systems that use a long range attraction, mediated by the three-dimensional (3D) particles, to create pairs of two-dimensional (2D) particles.
Since they consider either one species \cite{Nishida2009,Midtgaard2016} or a density imbalance of spin species \cite{Wu2016}, the exchange symmetry can not be satisfied by the singlet states in the $l = 0$ orbital mode and is forced to be in one of the spin triplet states of the $l = 1$ orbital mode.
Okamoto {\it et al.} \cite{Okamoto2017a} studied a system with an equal density of spin-half fermion species in the 2D plane, where the exchange repulsion does not prevent $s$-wave pairing.
They went beyond the mean-field analysis of \cite{Nishida2009,Wu2016,Midtgaard2016} by doing a functional renormalization-group study, and found a full range of orbital paring states from $s$-wave to $g$-wave paring.
Surprisingly, they found $p$-wave pairing occupied a minuscule portion of the 2D fermion phase diagram.
Furthermore, our analysis below suggests that this portion of the phase diagram would have small transition temperatures due to weak effective interactions.
This suggests that either the Fermi-repulsion is required to stabilize the $p$-wave phase in mixed-dimensions or that fluctuations destabilize the phase.

In this paper we address this question and find that, in fact, this system can be used to produce a robust $p$-wave pairing phase for 2D fermions, and that relatively large transition temperatures, of the order of 10\% of the bandwidth of the 2D fermions, can be achieved.
This is done by appropriately exploiting the long-range mediated interaction in a parameter space not considered in  previous works \cite{Okamoto2017a}.
We consider a 2D-3D mixture where the 3D particles mediate a long-range interaction between the 2D particles, which we take to be SU(2) symmetric fermions.
We study the phase diagram of the 2D fermions  and show that the interplay  between a repulsive contact interaction and a long-range 3D-mediated interaction can lead to a triplet $p$-wave pairing phase in a wide range of parameter space, and that it is stable to fluctuations.
If realized, this system would provide a highly controllable, weak-coupling analogue of superfluid $^3$He, where one could explore the full range of spin textures and triplet superfluid properties.

We begin this article by reviewing the physics of mediated interactions in mixed-dimension systems and demonstrate the ability to construct an arbitrary potential given many mediator types (Section~\ref{sec:mediators}).
In Section~\ref{sec:aRG} we review the functional renormalization group (fRG) approach and analytically argue why the $p$-wave phase is stable in this system.
In Section~\ref{sec:nfRG} we present numerical results that demonstrate the stability of the $p$-wave phase and highlight the nearby $s$-wave and $d$-wave instabilities. The $p$-wave pairing phase appears in a large region of the parameter space that has not been considered by previous authors studying this mixed-dimensional system.
Finally, we conclude with a discussion on experimental feasibility and the various $p$-wave pairing states and vortices that may potentially be studied in this system.

\section{2D Interactions via 3D Mediators}
\label{sec:mediators}

The full range of pairing states proposed in \cite{Okamoto2017a} and the possibility of $p$-wave pairing in mixed-dimensional cold atom systems is a product of the tunability of the long-range interactions.
This tunability comes from the ability to change the properties of the higher dimension particles in order to change the  mediated long-range interaction between the lower dimension particles.
In this article we consider 3D particles ($\phi_{r}$) which can either be bosons ($b_{r}$) or fermions ($f_r$).
The 3D particles mediate interactions between 2D fermionic particles($\psi_{r,\sigma}$) with spin $\sigma$.
The mediated interaction is computed in \cite{Okamoto2017a} and \cite{Nishida2009} for fermions and in \cite{Wu2016,Midtgaard2016} for bosons as the 3D particles.
The derivation is repeated here to demonstrate the control over the real space interaction for the 2D fermions and to highlight the similarity between various types of 3D mediators.

The action can be broken into the 2D part, the 3D part, and the interaction between the 2D and 3D particles:
\begin{eqnarray}
    S= S_{2}(\psi) + S_{3}(\phi) + S_{I}(\phi,\psi)
\end{eqnarray}
We take the interaction between the two types of particles as an s-wave contact interaction:
\begin{eqnarray}
    S_{I}(\phi,\psi) = g\sum_{r}\psi_{r}^{\dagger}\psi_{r}\phi_{r}^{\dagger}\phi_{r}
\end{eqnarray}
The strength and sign of the coupling can be tuned via Feshbach resonance.
When Fourier transformed, momentum is only conserved in the 2D plane:
\begin{eqnarray}
S_{I}(\phi,\psi) \!\!= \!\!
\frac{g}{V\beta}\!\!\sum_{\vec{k}',\vec{k},\vec{q},\vec{q}',q_z,q_z'}\!\!\!\!\!\!\!\!\!\!\!\psi_{\vec{k}'}^{\dagger}\psi_{\vec{k}}\phi_{\vec{q}',q_z'}^{\dagger}\phi_{\vec{q},q_z}\!\delta(\vec{k}'\!\!-\!\vec{k}\!+\!\vec{q'}\!\!-\!\vec{q})
\end{eqnarray}
where we have assumed the 2D plane to be located at the center of the 3D trap at $z=0$, the vectors ($\vec{k}$,$\vec{q}$\ldots) are 2D vectors, and the sums over Matsubara frequencies and spins are implicit.

In addition to the 2D and 3D traps for the two atom species, we study a system with an added 2D and 3D periodic optical lattice potential.
This allows us to slow the Fermi-velocity of the 2D lattice fermions so the interactions mediated by the 3D particles can be approximated as instantaneous.
The action of the 2D fermions is written as:
\begin{align}
    S_{2}(\psi) = & \sum_{\vec{k},\omega_{n}, \sigma} \psi_{\vec{k},n,\sigma}^{\dagger}(\varepsilon_{\vec{k}}-i\omega_{n})\psi_{\vec{k},n,\sigma}
    \nonumber \\
    & + \frac{U_{2}}{A \beta }\sum_{\vec{q},\vec{k_1},\vec{k_2}}\psi_{\vec{k_1}+\vec{q}}^{\dagger}\psi_{\vec{k_2}-\vec{q}}^{\dagger}\psi_{\vec{k_2}}\psi_{\vec{k_1}}
    \label{eq:s2}
\end{align}
\noindent where  $\varepsilon_{\vec{k}}=2t_{2}[\cos(k_{x})+\cos(k_{y})]$ is the lattice dispersion, $A$ is the area of the 2D lattice, and $\omega_{n}$ are fermionic Matsubara frequencies.
The coupling $U_2$ parameterizes a contact interaction between the 2D fermions and  depends on the depth of the 2D lattice.

Integration of the 3D particles modifies the chemical potential of the 2D fermions and generates an additional effective interaction.
Thus, from now on, the 2D chemical potential, $\mu_{2}$, will be implicitly understood to contain the modification by the 3D particles.
We first integrate the 3D particles as fermions ($\phi_{r}=f_{r}$) in a 3D lattice with the action:
\begin{align}
    S_{3} &= \sum_{\vec{k},k_z,\nu_{n}} f_{\vec{k},k_z,n,}^{\dagger} f_{\vec{k},k_z,n} G_{3}^{-1}(\vec{k},k_z,\nu_{n})
\end{align}
\noindent where $G_{3}(\vec{k},k_z,\nu_{n}) = [\varepsilon_{\vec{k},k_z} - i\nu_{n}]^{-1}$ is the propagator for the 3D particles, $\nu_{n}$ are the Matsubara frequencies for the 3D particles, and $\varepsilon_{\vec{k},k_{z}}=2t_{3}\left[ \cos(k_x)+\cos(k_y) + \cos(k_z) \right] -\mu_{3}$ is the 3D lattice dispersion.

The integration is done perturbatively and yields a one-loop particle-hole diagram for the effective interaction:
\begin{align}
    V_{f}(\vec{q},\omega_{n}) =\! \frac{-g^2A}{V^2\beta}
    \!\!\!
    \sum_{\vec{k},k_{z},k_{z}',\nu_{n}}
    \!\!\!\!\!\!\!\!
    G_{3}(\vec{k},k_{z},\nu_{n})G_{3}(\vec{k}\!+\!\vec{q},k_{z}',\nu_{n}\!+\!\omega_{n}\!)
    \label{eq:ferm_eff_1}
\end{align}
Converting to the continuum and integrating the Matsubara sum, we are left with:
\begin{align}
    \frac{-g^2}{(2\pi)^{4}}\!\int \!\! d\vec{k}dk_{z}dk_{z'} \frac{n_{f}(\varepsilon_{\vec{q},q_{z}}) - n_{f}(\varepsilon_{\vec{q}+\vec{k},q_{z}'}-i\omega_{n})}{\varepsilon_{\vec{q}+\vec{k},q_{z}'}-\varepsilon_{\vec{q},q_{z}}-i\omega_{n}} \ .
    \label{fermion_eff}
\end{align}
In order to have an effectively instantaneous interaction between the 2D particles mediated by 3D particles, we impose that $t_{2}\ll t_{3}$ ($\alpha \equiv t_3/t_2 \gg 1$), that is, the 3D particles move much faster than the 2D particles. We estimate that $\alpha$ can range from 3 to 190: where at the lower end, we used two similar mass particles, such as Li and Na, in lattices with similar depths;
while at the upper end, we are considering two particles with significantly different masses, such as Cs and Li, and that the lattice depth of the 2D particles decreases the tunnelling rate by a factor 10 \cite{Jaksch1998}.
The magnitude of the mediated interaction is of order $g^2/6 t_3$, and the integration over the 3D particles is done perturbatively, requiring $g < t_3$. In the next section, we perform a weak-coupling renormalization-group treatment of the 2D fermions interacting with the contact interaction $U_2$ and the mediated interaction $V_f$. This requires the overall 2D interaction to be smaller than half the bandwidth of the 2D fermions.
If we only take the mediated interactions into account, its ratio with the bandwidth of the 2D particles is 
$O(g^2/4t_{2}6t_{3})=O(\alpha g^2/24t_{3}^2)$. 
Thus, at first glance, it appears we need $\frac{g}{2\sqrt{6}t_{3}}<1/\sqrt{\alpha}$ for the renormalization-group analysis for the 2D fermions to be valid. We will return to this point later and see that this stringent condition can be avoided.

\begin{figure}
    \includegraphics{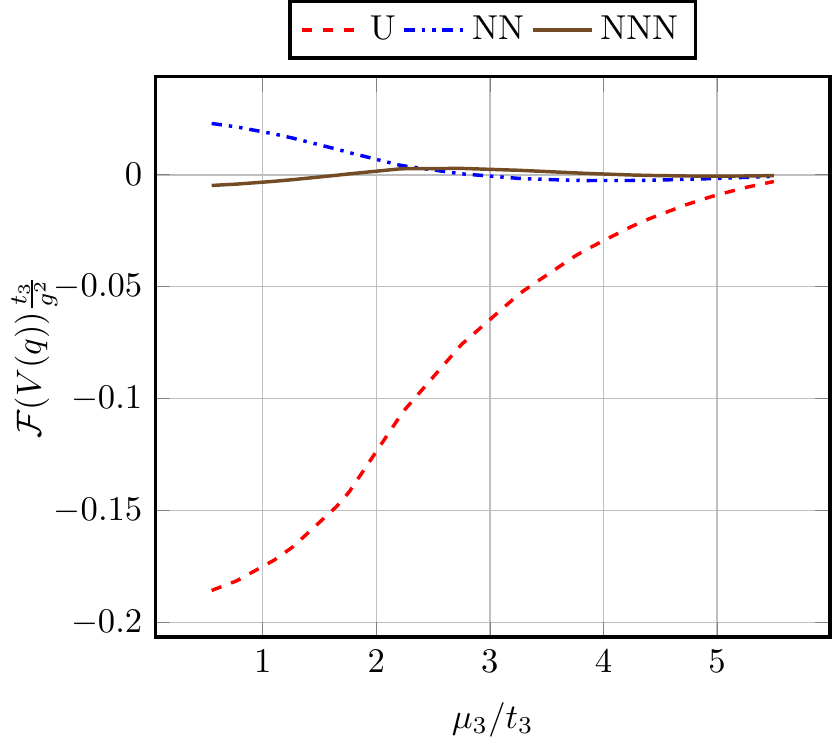}
    \caption{This is a plot of the Fourier transform at various distances for a 2D mediated interaction mediated by 3D lattice fermions. U is the on-site interaction strength, NN is for the nearest-neighbour and NNN is for the next-nearest-neighbour. The y-axis is in units of $t_{2}$ and the x-axis is in units of $t_{3}$. }
    \label{fermion_V_r}
\end{figure}

Despite the lattice dispersion preventing an analytic calculation of Eq. \ref{fermion_eff}, we ignore retardation effects by setting $\omega_{n}=0$ and numerically integrate Eq.~\ref{fermion_eff}.
The Fourier transform of $V_f(\vec{q},\omega_n=0)$ for various distances (on-site, nearest-neighbor, next-nearest-neighbor) is plotted in Fig.~\ref{fermion_V_r} as a function of the 3D chemical potential. The ability to tune the 3D particle density, determined by $\mu_3$, can thus be viewed as a way to tune the relative values and signs of the different components of the mediated interaction. A key thing to note in the mediated interaction (Fig.~\ref{fermion_V_r}) are the Friedel oscillations, where the sign of the interaction changes at different distances, and the nearest-neighbor interaction is exponentially smaller than the on-site interaction.
This means that, if the on-site (non-mediated) contact interaction $U_2$ is strong enough to cancel the on-site mediated interaction, an integration of the effective 2D fermionic action done during the renormalization-group analysis (next section) will still be perturbative when $g/2\sqrt{6}t_{3}>1/\sqrt{\alpha}$.
Thus, while Fig.~\ref{fermion_V_r} depicts a negligible nearest-neighbor interaction, it can still be formidable, scaling with $\alpha$.

This calculation was done for a finite-sized lattice in~\cite{Okamoto2017a} and a similar calculation was done by~\cite{Nishida2009} where the lattice potential is not included.
Without the lattice, there is no natural ultraviolet cutoff and regularization is needed to deal with an ultraviolet divergence.
The end result is effectively the same as when the lattice spacing is much smaller than the inter-particle spacing, i.e.\ at low densities.
The difference is at higher densities, where, in the system without the lattice potential, there is no interplay between the lattice spacing and the average particle spacing.
For the lattice system, the interaction strength will peak at half filling where the primary mediator switches to a lattice hole and the hole density decreases until the band is full and no longer mediates interactions.
While for the system without the lattice potential, the density of states of the scattering mediators is always increasing with the area of the Fermi-surface.

For bosonic mediators, we assume the 3D bosonic system has condensed to form a superfluid and interactions are mediated by Bogoliubov particles with dispersion $\omega_{q}^2=\varepsilon_{0}(q)\left[\varepsilon_{0}(q)+\Delta\right]$.
Where $\varepsilon_{0}$ is the dispersion for the free bosons and $\Delta$ is the superfluid gap.
The interaction between the 2D fermions and the Bogoliubov quasi-particles is written:
\begin{align}
    S_{I}(b,\psi) =&
    \frac{g}{V\beta}& \sum_{\vec{k},\vec{q},q_{z}}\psi^{\dagger}_{\vec{k}+\vec{q}}\psi_{\vec{k}} \sqrt{\frac{N_b \varepsilon_{0}(\vec{q},q_z)}{\omega_{\vec{q},q_z}}}(a^{\dagger}_{-\vec{q},-q_z}+a_{\vec{q},q_z})
\end{align}
\noindent where $a$ and $a^{\dagger}$ are the creation and annihilation operators for the Bogoliubov quasi-particles, and $N_b$ is number of condensed bosons.
The action for the quasi particles is now quadratic and can be integrated exactly yielding an effective interaction:
\begin{align}
    V_{b}(\vec{q},\nu)=-\frac{g^2N_{0}}{2\pi V}\int dq_{z} \frac{\varepsilon_{0}(\vec{q},q_z)}{\omega_{\vec{q},q_z}^{2}+\nu^{2}} \ .
\end{align}
Fourier-transforming to real space, we get:
\begin{align}
    V_{b}(\delta \vec{r},\nu)=-\frac{g^2n_{0}}{(2\pi)^3}\int dq_{z}d\vec{q} e^{i\vec{q}\cdot\delta\vec{r}}\frac{\varepsilon_{0}(\vec{q},q_z)}{\omega_{\vec{q},q_z}^{2}+\nu^{2}}
    \label{eq:bosonmed}
\end{align}
where $\delta \vec{r}$ is the distance between the two particles and $n_{0}$ is the boson density.
This interaction is the same as for 3D bosons mediating other 3D particles but with $\delta r_{z}=0$.
Again, we assume the 3D particles move much faster than the 2D particles and  the interaction potential becomes of the Yukawa form, with a mass equal to the condensation gap $\Delta$.
Thus, with regards to the sign and relative strength of the on-site and nearest-neighbor interactions, the bosonic mediator is the same as the fermionic mediator at low 3D densities.
This means that, in terms of the phase diagram for low-density 3D fermions derived from this effective action, free-fermion and free-boson mediators are effectively the same.
This does not mean there are no great physical differences between these systems.
These differences can have great importance related to the feasibility of an experiment.
For example, mixed-dimensional bosonic systems have been shown to suffer losses from three-body Effimov physics \cite{Nishida2008a} and free fermions can have much larger Fermi-velocities than their low-density lattice counterparts.

We reproduce these calculations here to demonstrate the ability to arbitrarily control the effective on-site and nearest-neighbor interactions.
For a single mediator, the knobs are the coupling strengths $g$ and $U_2$, the mediator density via $\mu_{3}$, and the tunnelling rates, $1/t_2$ and $1/t_3$.
Importantly, for lattice fermions in 3D, $\mu_{3}$ allows one to choose the sign of the nearest-neighbor interaction.
The tunneling rate $1/t_3$ controls the overall strength of the mediated interaction while the on-site coupling $U_2$ can adjust the overall on-site interaction strength to a desired value. In particular, the on-site interaction can be made to vanish, or to be purely repulsive, thus suppressing $s$-wave BCS pairing of the 2D fermions. 

While experimentally infeasible, it is entertaining to note that this process  could theoretically be extended to arbitrary control of the effective interaction strength of the $n^{th}$-nearest-neighbor sites.
In this generalization, there are $n$ free parameters from the $n$ 3D particle densities which can be used to tune the interaction at $n$ different interaction distances.
Then, the on-site coupling strength, $U$,  can correct whatever on-site interaction is left over.

\section{Renormalization Group Analysis of Triplet Pairing Instability}
\label{sec:aRG}

The versatility of the 2D interaction via 3D mediators described in the previous section gives a wide range of control over the pairing instability of the 2D fermions.
In this section, we demonstrate that a robust pairing instability of the triplet $p$-wave type can be created that dominates a significant region of parameter space.
To this end, we will work with an explicitly 2D action in the weak coupling limit and consider fluctuations directly via the functional Renormalization Group (fRG) \cite{Shankar1994,Metzner2012}. fRG is a broad class of RG schemes that specifies the dependence of some functional on some parameter via the flow equations.
When this functional is the effective action at a given energy scale, one can obtain Wilson-like RG equations for the various $n$-point functions~ \cite{Metzner2012}.
Here we consider a 2D effective action for fermion with (pseudo-)spin.
In a perturbative expansion one can focus on the $n$-point functions with small $n$.
We write the effective action for the 2D fermions as:
\begin{align}
    \Gamma^{\Lambda}(\psi) = \sum_{i}\Gamma_{2}^{\Lambda}\psi_{i}^{\dagger}\psi_{i}+\sum_{i,i',j,j'}\Gamma_{4}^{\Lambda}(i,i',j,j')\psi_{i'}^{\dagger} \psi_{j'}^{\dagger}\psi_{i}\psi_{j}
\end{align}
where $i$, $j$, etc. are field variables that carry momentum, spin and Matsubara frequency: $(\vec{k_{i}},\omega_{n_{i}},\sigma_{i})$.
For the initial conditions at $\Lambda=\Lambda_{0}$, the effective action is given as the bare 2D action (Eq.~\ref{eq:s2}) with a modified interaction, $\Gamma_{4}^{\Lambda_0}=[U_{2}+V_{f}(\vec{q},0)]/(A\beta)$,
where $V_{f}(\vec{q},0)$ and $U_{2}$ are defined in Eq.~\ref{eq:s2} and Eq.~\ref{eq:ferm_eff_1}, respectively.
We will discuss the generalization to other 3D mediators at the end of this section.

\begin{figure}[t!]
    \captionsetup[subfigure]{labelformat=empty}
    \begin{subfigure}[h]{0.15\textwidth}
        \centering
        \includegraphics[width=1\textwidth]{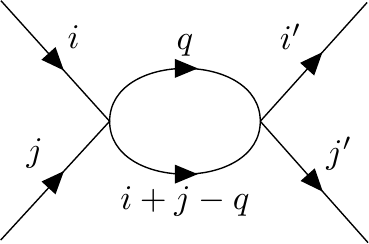}
        \caption{PP}
    \end{subfigure}
    ~
    \begin{subfigure}[h]{0.15\textwidth}
        \centering
        \includegraphics[width=1\textwidth]{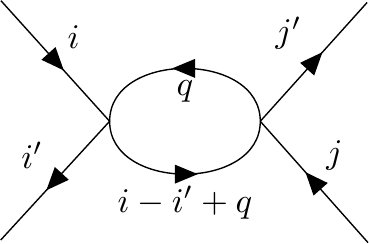}
        \caption{PH}
    \end{subfigure}
    ~
    \begin{subfigure}[h]{0.15\textwidth}
        \centering
        \includegraphics[width=1\textwidth]{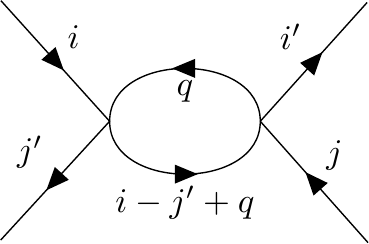}
        \caption{PH'}
    \end{subfigure}

    \caption{The four-point beta functions. The PP diagram is responsible for pairing instabilities while the PH and PH' diagrams are responsible for density wave instabilities.}
    \label{fig:fpbf}
\end{figure}

Our analysis on the relevant couplings focuses on the flow of two-particle scattering or the four-point function, $\Gamma_{4}^{\Lambda}$.
At one-loop order, one can ignore the flow of the two-point function when considering the flow equations for the four-point function.
The flow of the four-point function, $\Gamma_{4}$, contains contributions from the three diagrams shown in Fig.~\ref{fig:fpbf}.
The integral for $\beta_{PP}$ (defined in Fig.\ref{fig:fpbf} is written:
\begin{align}
    \beta_{PP}&(\Gamma_{4}^{\Lambda}(i,i',j,j'))=
    \\ \nonumber
    &\frac{1}{2}\sum_{q}\Gamma_{4}^{\Lambda}(i,q,j,i+j-q)\Gamma_{4}^{\Lambda}(q,i',i+j-q,j')
    \\ \nonumber
    &\times \frac{[\Theta(\varepsilon_q'-\Lambda)\delta(\varepsilon_{q}-\Lambda) + q \leftrightarrow q']}{(\varepsilon_{q}-i q_{0})(\varepsilon_{q'}-iq'_{0})}
    \label{ppbeta}
\end{align}
where $q'=i+j-q$ and $\Theta$ is the step function.
All three beta functions have two internal legs.
One internal leg is restricted to an equal-energy surface at the cutoff, while the other is restricted to the momentum above the cutoff and determined by momentum conservation.
The magnitudes of the beta functions are inversely proportional to the energy of the two internal legs.
Therefore, the strongest flowing diagrams will have the momentum-conserved leg close to the cutoff throughout integration. 
This condition for the PP diagram is then:
\begin{eqnarray}
    \varepsilon_{i+j-q}=\Lambda
    \label{PP}
\end{eqnarray}
for all $q$ on the integration surface.
Similarly, the condition for the PH contribution is
\begin{eqnarray}
    \varepsilon_{i-i'+q}=\Lambda \ .
    \label{PH}
\end{eqnarray}
for all $q$ on the integration surface.

Furthermore, couplings with external legs on the Fermi-surface will always be flowing, while couplings above the cutoff will not.
Therefore, the couplings on the Fermi-surface that satisfy the above conditions will flow the strongest.
In fact it can be shown that all other couplings are irrelevant \cite{Shankar1994,Salmhofer1999}.
These conditions in 2D reduce the parameterization of the important scatterings to two angles and the spin dependence: $\Gamma_{PP}^{\Lambda}(\sigma,\theta_{i},\theta_{i'})$ and $\Gamma_{PH}^{\Lambda}(\sigma,\theta_{i},\theta_{j})$.
Given SU(2) symmetry the spin dependence breaks down into singlet and triplet scattering, $\sigma=S,T$.

When the effective chemical density of the 2D fermions, $\mu_{2}$, is not zero (system away from half-filling), the Fermi-surface scatterings do not satisfy Eq.~\ref{PH}.
On the other hand, the Fermi-surface at all fillings obeys inversion symmetry, and the PP condition, Eq.~\ref{PP}, is always satisfied for couplings with $i=-j$.
Therefore, the following couplings are always flowing: $\Gamma_{PP}^{\Lambda}(\theta_{i},\theta_{i'})$.
When $\mu_{2}$ is close to the bandwidth, the energy surfaces close to the Fermi-surface are circular and the frequency integral in Eq.~\ref{ppbeta} can be evaluated at zero temperature directly yielding \cite{Shankar1994}:
\begin{align}
    \Lambda\frac{d}{d\Lambda}\Gamma_{PP}^{\Lambda}(\theta_{i},\theta_{i'}\!)= \!\frac{1}{8\pi^{2}}\!\! \int\! \frac{d\theta}{2\pi} \Gamma_{PP}^{\Lambda}(\theta_{i},\theta)\Gamma_{PP}^{\Lambda}(\theta,\theta_{i'}) \ .
\end{align}
This equation can be decomposed into angular momentum modes, $l$, and the resulting set of differential equations solved by:
\begin{align}
    \Gamma_{l}^{\Lambda} = \frac{\Gamma_{l}^{\Lambda_0}}{1-\frac{\Gamma_{l}^{\Lambda_0}}{4\pi}\ln(\frac{\Lambda}{\Lambda_0})}
    \label{ppsol}
\end{align}
\noindent where, $\Gamma_{l}^{\Lambda_0}$ is the decomposition of the PP four-point function, $\Gamma_{PP}^{\Lambda}$, at the scale of the bandwidth, $\Lambda=\Lambda_0$.
When $\Gamma_{l}^{\Lambda_0}$ is attractive, Eq.~\ref{ppsol} has a divergence at a critical scale,
\begin{align}
    \Lambda_{c}=\Lambda_{0} \exp{\left(-\left|\frac{4\pi}{\Gamma_{l}^{\Lambda_0}}\right|\right)}
    \label{ppsd}
\end{align}
indicating that the system flowed to a new fixed point.
For finite temperatures, the singularity in Eq.~\ref{ppbeta} is smoothed and for temperatures above the critical scale, the couplings stay near the free (Fermi-liquid) fixed point.
For temperatures below the critical scale, the couplings diverge and the flow can not be continued to the new fixed point. 
A proper mean-field analysis can be used to determine the properties of the new fixed point.
For the PP divergences, mean field suggests a BCS pairing state describes the new fixed point, where the angular momentum and the spin state of pairs are determined by the leading $\Gamma_{l}^{\Lambda}$.

Eq.~\ref{ppsol} indicates the leading divergence will correspond to the attractive mode with largest magnitude of $\Gamma_{l}^{\Lambda_0}$.
Therefore, we will have triplet pairing when the largest $\Gamma_{l}^{\Lambda_0}$ has an odd $l$.
On-site, local interactions have a flat Fourier transform and thus only contribute to the $s$-wave interaction.
Nearest-neighbor interactions, on the other hand, involve four sites that modify the rotational symmetry and therefore can contribute to the $l=1$ ($p$-wave) or $l=2$ ($d$-wave) modes.
So, given the tunability of 3D mediators, triplet pairing can be induced by an attractive, long-range, effective action.
The conditions for $p$-wave ($l = 1$) pairing in terms of $\Gamma_{l}^{\Lambda_0}$ are: $\Gamma_{1}^{\Lambda_0}<0$ and $|\Gamma_{1}^{\Lambda_0}|>|\Gamma_{l}^{\Lambda_0}|$, for $l\ne 1$.

This requires that the nearest-neighbor attraction be stronger than the on-site attraction, otherwise $s$-wave pairing will dominate.
In spin-less or spin-imbalanced systems, the exchange repulsion guarantees that this is the case.
For $SU(2)$ symmetric systems, the mediated on-site attraction, found in Fig. \ref{fermion_V_r} and Eq.~\ref{eq:bosonmed}, needs to be compensated by a repulsive contact interaction $U_2$ between the 2D particles.

\begin{figure}
    \includegraphics{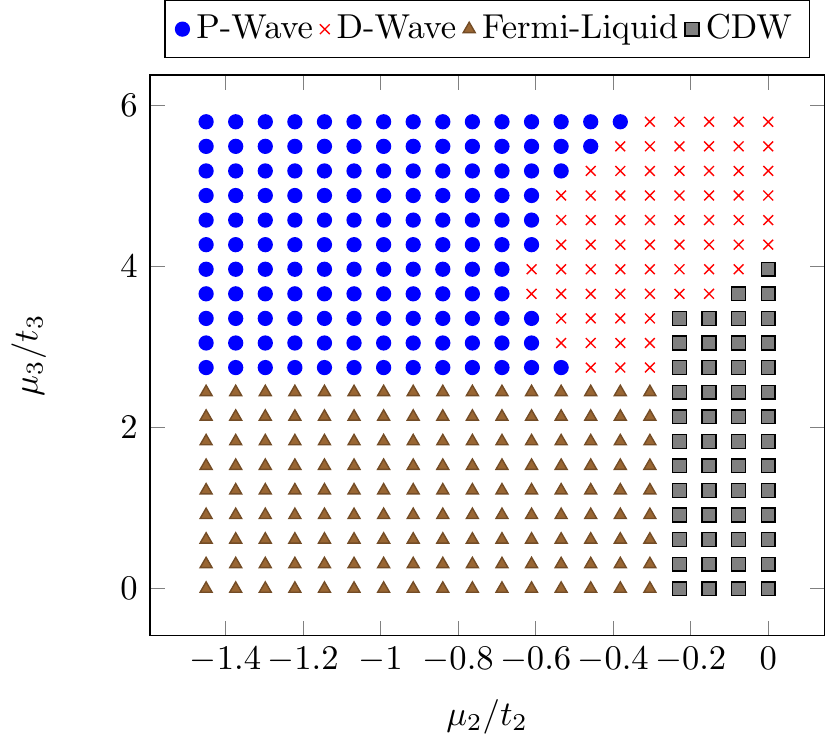}
    \caption{Phase diagram suggested by bare interaction. This is for fixed $g$ and for $U_2$ set to cancel the on-site interaction.}
    \label{ipd}
\end{figure}
This compensation still leaves the possibility of $d$-wave pairing and we check this numerically.
This calculation begins by breaking the two angles in the particle-particle couplings, $\Gamma_{PP}^{\Lambda_0}(\sigma,\theta_{i},\theta_{i'})$, into 16 patches.
We then compute the eigenvectors and eigenvalues of the matrices $\Gamma_{PP}^{\Lambda_0}(\sigma,i,i')$.
The number of zeros in the eigenvector determine the angular momentum $l$.
To satisfy the $2\pi$ periodic boundary condition around the Fermi-surface, $\Gamma_{PP}^{\Lambda_0}(\sigma,i,i')$ must have eigenvectors with $2l$ zeros.
To satisfy exchange symmetry, matrices with $\sigma=T$ will have odd $l$, while those with $\sigma=S$ will have even $l$.
For energy levels away from the circular Fermi-surface, the decomposition of the beta functions will depend on the shape of the Fermi-surface,
 but the general feature that negative eigenvalues diverge at a given scale will remain.
This is because there is still inversion symmetry and the PP graph still flows.
This way we can still infer what fixed point the RG equations will flow to by decomposing the initial four-point function, $\Gamma_{PP}^{\Lambda_0}(\sigma,\theta_{i},\theta_{i'})$, in to its angular momentum components.

In search of the $l=1$ triplet mode, we consider the $\mu_{2}$-$\mu_{3}$ plane where we fix the on-site contact interaction, $U_2$, to cancel the on-site component of the mediated interaction($V_{q}$) such that $U_{eff}/t_2=1$, where $U_{eff}$ is the net on-site component of the effective interaction.
The overall strength of the mediated interaction will determine the critical scale but will not affect which pairing state appears.
For this, we set the overall strength of the contact interaction between the 3D and 2D gases to be $(\alpha g^2/t^2_{3}) = 100$.
The phases predicted by the decomposition of the initial four-point function is shown in Fig. \ref{ipd}.
For $\mu_{3}/t_{3}<2.2$, the nearest-neighbor interaction is also repulsive, so the only potentially divergent coupling flows to zero.
Thus, we predict that this part of the phase diagram will be a Fermi-liquid at all experimentally-relevant energy scales.
For $\mu_{3}/t_3>2.2$, the nearest-neighbor interaction is attractive and we expect pairing.
For $\mu_{2}/t_2<-0.6$, $p$-wave pairing dominates but for $\mu_{2}>-0.6$, the symmetry of the Fermi-surface favors $d$-wave pairing.
For $\mu_{2}/t_2 = 0$, the Fermi-surface is nested and we expect the PH couplings to diverge first and the system to favor a CDW.

We expect similar results for free fermionic and bosonic mediators.
Given that the on-site contact interaction cancels the on-site mediated interaction, a higher angular momentum pairing state will dominate at low temperatures.
Which pairing state dominates will  depend on the geometry of the Fermi-surface, and as the 2D fermions approach half-filling, the $p$-wave instability will give place to a $d$-wave one. 
The CDW will continue to persist at half-filling due to the nesting of the Fermi-surface.
The primary difference is that the bosons and fermions do not acquire a repulsive nearest-neighbor interaction and destabilize the $p$-wave phase at a specific doping.

Similar results are also expected to hold in the spinless systems studied with mean field by Nishida and Wu \cite{Nishida2009,Wu2016,Midtgaard2016}.
In this system, the effective interaction, $\Gamma_{4}^{\Lambda_0}$, is the same as the triplet component of the $SU(2)$ symmetric system.
The PP instability is still present, but the exchange symmetry only allows odd $l$.
Thus, the negative initial eigenvalue for $l=1$ will still be dominant, but it will not have to compete with $s$ or $d$-wave.
Furthermore, in the system with fermionic mediators on a lattice, the transition to a Fermi-liquid at higher densities will still occur as the $l=1$ initial eigenvalue changes sign.

\section{Numerical fRG Results}
\label{sec:nfRG}
The results above give our analytic expectation for when the 2D effective action flows to a triplet pairing fixed point.
In this section, we numerically solve the RG flow equations, with the intent to directly consider anisotropic effects of the Fermi-surface and the interaction between different pairing instabilities. 
To this end, we compute fRG flows for the Wick ordered effective action, $W^{\Lambda}(\psi)$ which generates Wick ordered $n$-point functions \cite{Metzner2012, Salmhofer1999,Halboth2000}.
This functional directly reproduces the effective action as $\Lambda \rightarrow 0$ and has the numerically appealing feature that its flow equations are local in the cutoff \cite{Halboth2000}.
Since the $\mathcal{W}^{\Lambda}$ only reproduces the effective interaction in the zero cutoff limit, we predict the low energy phases by tracking how correlation functions diverge as the cutoff is lowered.
This is because, above the critical temperature, $\mathcal{W}^{\Lambda}$ does not have any divergences and correctly gives the correlation functions.
As one approaches the critical temperature, critical fluctuations lead to divergences in the correlation functions.
This will again be captured by $\mathcal{W}^{\Lambda}$ and will continue to diverge in the same way as we reduce the temperature.
This is because we are still starting from the free fixed point. Therefore, we can use which correlation diverges first to determine the low energy phase \cite{Halboth2000}.
We consider particle-particle correlations to identify pairing phases:
\begin{align}
    \mathcal{X}_{PP}^{\Lambda}(f,s) = \left<\left|\int d\theta f(\theta)s_{\sigma,\sigma'} \psi_{\sigma}(k_{f},\theta) \psi_{\sigma'}(k_{f},-\theta)\right|^2 \right>^{\Lambda} \ ,
\end{align}
where the expectations are computed at a cutoff $\Lambda$ using $\mathcal{W}^{\Lambda}$, $f(\theta)$ determines the pairing symmetry, and $s_{\sigma,\sigma'}$ determines whether the system is in a singlet or triplet pairing state.
Similarly, particle-hole correlations are used to identify density-wave phases:
\begin{align}
    \mathcal{X}_{PH}^{\Lambda}(s) = \left<|\int dk s_{\sigma,\sigma'} \psi_{\sigma}^{\dagger}(\vec{k}) \psi_{\sigma'}(\vec{k}+\vec{q})|^2 \right>^{\Lambda} \ .
\end{align}
Here $s_{\sigma,\sigma'}$ determines if we are considering spin or charge density-waves.
At the beginning of the flow these correlations are of $O(1)$.
As the cutoff approaches the critical scale, the dominant correlation diverges, and we can approximate the critical scale for a given correlation as the scale when that correlation reaches some value, which here we choose to be of $O(100)$.

As is standard with numerical fRG calculations \cite{Halboth2000,Platt2013a,Zanchi2000,Huang2013,Bhongale2012}, we flow a finite number of couplings by projecting the momentum to a finite number of patches on the Fermi-surface.
These patches are identified by their angle, $\theta_{i}$ for $i \in (1,m)$.
Since given three momenta, the fourth momentum is specified by momentum conservation, there are in principle $m^{3}$ couplings and $m^{3}$ coupled differential equations to be solved for each $\Gamma^{\Lambda}(i,j,k)$.
Due to this computation complexity we work with $m=16$, and do not explore the full four-dimensional parameter space ($\mu_{2},\mu_{3},U,g$).
Instead, we focus on specific cuts to confirm the above picture and understand the stability of the triplet pairing phase.

\begin{figure}
 \includegraphics{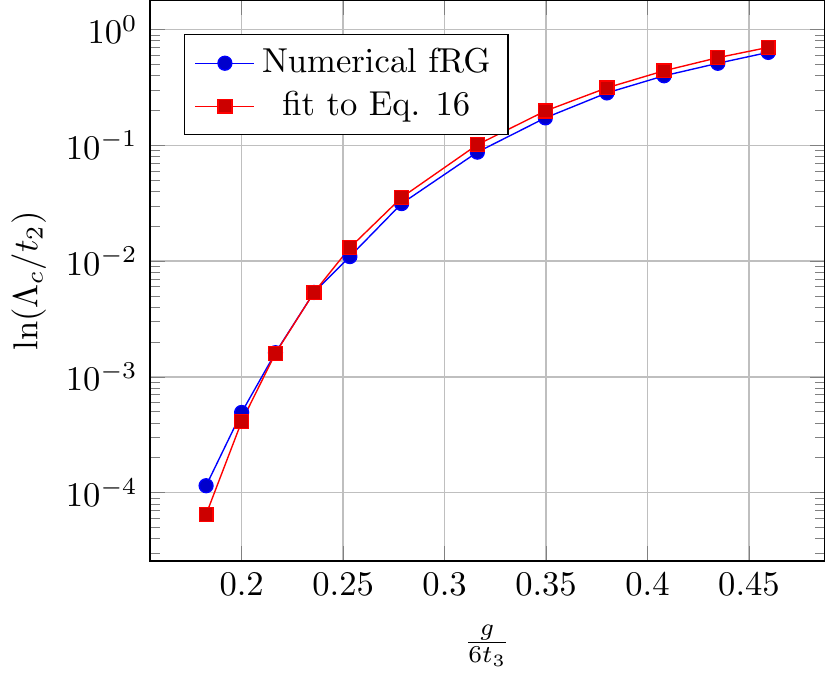}
    \caption{As we increase the mediated interaction, the critical scale rises exponentially until we are in the strong coupling limit.  The overall interaction strength depends on $\alpha$ which we set to 100 in this figure. }
    \label{csvV3}
\end{figure}
First, we test the estimation given by Eq.~\ref{ppsd}, which gives the triplet-pairing critical temperatures as a function of $1/|\Gamma_{l}^{\Lambda_0}|$.
To this end, we make the first cut deep within the expected triplet-pairing phase at $\mu_{2}/t_{2}=-1.4$ and $\mu_{3}/t_{3}=3.85$.
As before, we set the on-site contact interaction, $U_2$, to cancel the on-site component of the mediated interaction($V_{q}$) such that $U_{eff}/t_{2}=1$: suppressing $s$-wave pairing.
We confirm that the $p$-wave triplet $\mathcal{X}_{PP}^{\Lambda}(p,T)$ correlation diverged first, and in Fig. \ref{csvV3}, we plot the dependence of the critical scale on the expansion parameter of the 3D particle integration.
We then fit this to the prediction of Eq.~\ref{ppsd}.
$\Gamma_{l=1}^{\Lambda_0}$ in Eq.~\ref{ppsd} is the $l=1$ component of the effective interaction and is proportional to the microscopic parameters $g$, $t_3$, $\alpha$, $\Gamma_{l=1}^{\Lambda_0} = \gamma \alpha(g/t_{3})^2$.
Thus, the fit only has one free parameter $\gamma$.
Despite the non trivial geometry of the Fermi-surface, an aggressive patching scheme, and the inclusion of the particle-hole beta functions, the scaling is well described by Eq.~\ref{ppsd}.
We find that $\gamma=9\times 10^{-3}$. This compares well with our expectation that the $l=1$ mode is coming from the effective nearest-neighbor attraction(Fig \ref{fermion_V_r}), which is approximately $2\times 10^{-3}$ at $\mu_{3}/t_{3}=3.85$.
This corresponds to a nearest neighbor interaction strength of about half the bandwidth, $4t_{2}$.

At large values of $g/6t_{3}$, the perturbative integration of 3D particles giving the mediated interaction may be called into question.
This can be resolved by increasing $\alpha$, thus allowing the effective nearest-neighbor strength to be large, while keeping $g/6t_{3}$ still in the perturbative regime.
There is still a question on the validity of the one-loop fRG equations, because the effective nearest-neighbor interaction still needs to be sufficiently large
for the critical temperature to be about 10\% of the bandwidth in order to be experimentally accessible.
This interplay between a high critical temperature and perturbative validity will be discussed further in the Section~\ref{sec:discussion}.

\begin{figure}
    \includegraphics{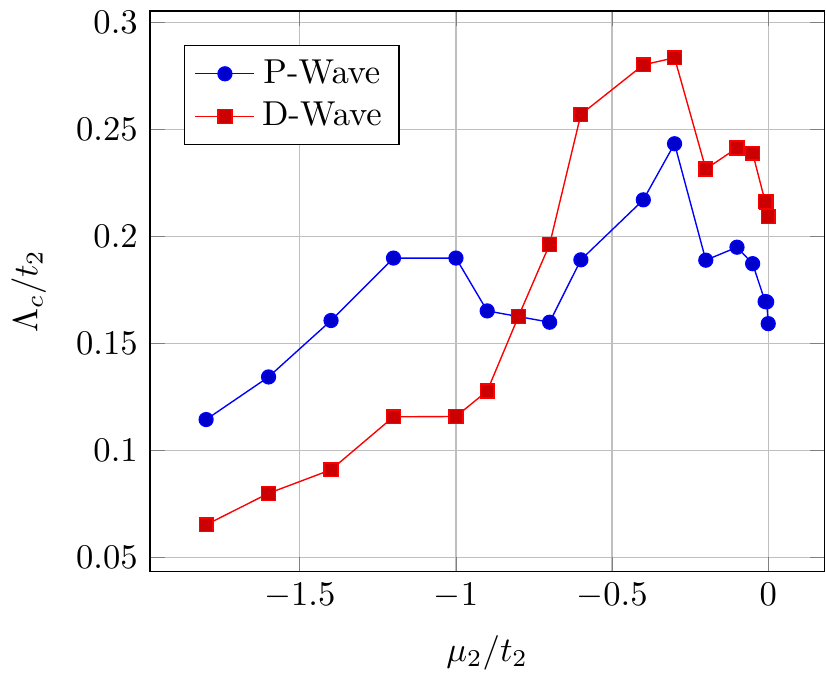}
    \caption{This plot depicts the instability of $p$-wave pairing to $d$-wave pairing upon varying the 2D particle density.  This cut is at $\mu_{3}/t_{3} = 3.85$, $U_{eff}/t_{2}=1$, $\alpha=150$ and $g/t_{3}=.86$}
    \label{mu2I}
\end{figure}

In the previous section, we predicted that as we decreased the doping of the 2D fermions, we would transition from triplet pairing to $d$-wave singlet pairing.
This transition was confirmed numerically and is depicted in Fig.~\ref{mu2I}. 
This figure shows the transition to $d$-wave at $\mu_{2}/t_{2}=-0.8$.
Away from the $p$-wave to $d$-wave transition, the difference between the critical scale of $p$-wave and that of the $d$-wave is independent of $\mu_{2}$.
This is a failure of perturbative fRG equations: when the scatterings responsible for the leading correlation function start to diverge, they also drive the divergence of non-leading correlation functions. 
Therefore, the non-leading correlations lag behind the leading correlation in a way independent of $\mu_2$.
It is also worth pointing out that we do not see any exponential scaling in the $p$-wave interaction because the relative strength of the $l=1$ interaction does not change as we vary $\mu_{2}$.

\begin{figure}
    \includegraphics{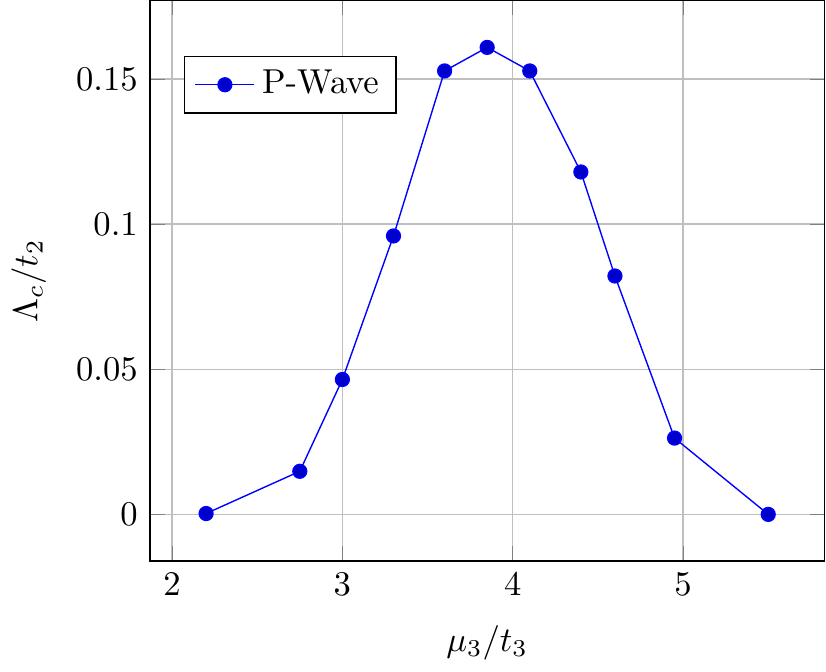}
    \caption{This plot depicts the dependence of the critical temperature of $p$-wave pairing on the 3D particle density.  This cut as the same interaction parameters as in Fig. \ref{mu2I}: $U_{eff}/t_{2}=1$, $\alpha=150$ and $g/t_{3}=.86$}
\label{fig:mu3I}
\end{figure}

Next, we consider doping in the 3D lattice fermions with a cut at $\mu_{2}/t_{2}=-1.4$.
Our results are plotted in Fig.~\ref{fig:mu3I} and confirm our expectations from the previous section.
At low 3D particle density, there are few particles to mediate interactions and the system remains a Fermi-liquid to lower temperatures.
As the 3D particles approach half filling, the nearest-neighbour interaction becomes repulsive and the $p$-wave instability disappears.
While at intermediate doping $\mu_{3}=3.85$ the $p$-wave instability reaches its maximum value.

\begin{figure}
    \includegraphics{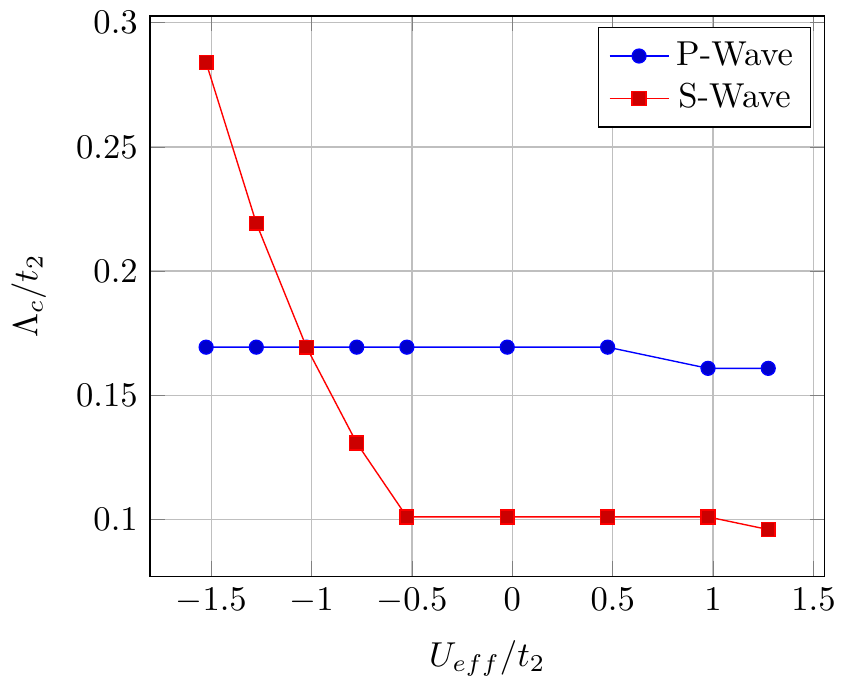}
    \caption{As we reduce the on-site contact interaction, the effective on-site interaction becomes negative and $s$-wave pairing starts to dominate.  This cut is at $\mu_{3}/t_{3} = 3.85$, $\mu_{2}/t_{2}=-1.4$, $\alpha=150$ and $g/t_{3}=.86$ }
    \label{UI}
\end{figure}
Finally, we consider the stability of the $p$-wave phase to the strength of the on-site contact interaction.
In Sec.~\ref{sec:aRG}, we pointed out that it is required to be strong enough to cancel the on-site attraction due to the mediated particles. 
In the previous cuts we always set the contact interaction so the effective on-site interaction was $U_{eff}/t=1$.
In Fig.~\ref{UI},we plot a cut where all parameters are fixed at their optimal values for $p$-wave pairing, and we vary the on-site contact interaction.
Once the effective on-site interaction becomes attractive, the $s$-wave critical scale begins to grow until it overcomes the $p$-wave pairing.
The transition point happens below 0 because the $l=1$ component still dominates until this point.
As we decrease the $l=1$ component by changing $\mu_{3}$ or reducing $g/t_{3}$, the transition point will increase to 0.

\section{Discussion}
\label{sec:discussion}
A major experimental barrier to realizing this system is cooling to achieve the predicted transition temperatures.
In the weak-coupling limit, fRG analysis clearly demonstrates that, at low enough temperatures, the Fermi-liquid phase is unstable to triplet-pairing.
Furthermore, if the nearest-neighbor interaction is close to, but less than, half of the bandwidth, the critical temperature will be about $10\%$ of the bandwidth (an experimentally accessible temperature).
There were two perturbation arguments used in reaching this result and we must be careful that they are not invalidated at the experimentally interesting limit.
The first is the perturbation theory used to derive the mediated interaction.
We can insure that $g/6t_{3}$ is small while still having a strong nearest-neighbour attraction by increasing the ratio $\alpha \equiv t_3/t_2$, i.e.\ using particles with significantly different masses and increasing the lattice depth of the 2D particles.
The second pertubative argument we used was in the expansion used to derive the $\beta$ functions.
Here we require the effective 2D interaction at the initial scale is smaller than the bandwidth.
Fortunately, the validity of perturbative-fRG has been shown \cite{Shankar1994} to extend into this limit.
The argument is based on kinematic constraints, similar to those that derived Eq.~\ref{PP} and Eq.~\ref{PH}, which suppress the integration of internal legs in higher order diagrams.
A direction that goes beyond the scope of this paper is to push $g/t_{3}$ beyond the perturbative limit.
In this case, one still expects interspecies interactions to mediate some form of long-range attraction and a mean-field analysis in similar systems has suggested a triplet instability \cite{Nishida2009,Wu2016,Midtgaard2016}.
Thus, despite possible experimental challenges with cooling, our results suggest that fluctuations are compatible with the mean-field analysis of previous studies and this system has a robust $p$-wave pairing instability in a large portion of its phase diagram.

This work goes beyond previous studies of $p$-wave pairing in mixed-dimension by demonstrating a mixed-dimension system with $SU(2)$ (pseudo)spin symmetric fermions can be unstable to triplet pairing.
Therefore, if implemented, this system could be used as a test bed for a wider range of triplet pairing physics.

The simplest example of a triplet-pairing-specific phenomenon is that the (pseudo)paramagnetic response can remain finite at zero temperature due to the degeneracy of the state to rotations in the (pseudo)spin degree of freedom \cite{Leggett1975}.
Similar to the spin-less case, the $p$-wave will favour a $p_x+ip_y$ order parameter, as opposed to a strictly $p_x$ or $p_y$.
This means the 2D fermion pairing-state will resemble the pairing state in $Sr_2RuO_4$ \cite{Maeno2012,Kallin2016}, which is expected to be a chiral superconductor and host Majorana fermions in its vortices.
This would also resemble the A phase in liquid $^3$He, but with the pairs restricted to two-dimensions and the orbital angular momentum pointing out of plane.
Therefore, by adding a couple layers to the 2D lattice, the orbital angular momentum could rotate into the 2D plane, and it may be possible to observe topological defects like the Shankar monopole \cite{Shankar1977a, Nakahara1987}.

In this work, we have studied 2D fermions that have a long-range attraction which can be mediated by three different types of 3D particles.
Focusing on 3D fermions in a lattice, we have shown that a triplet, $p$-wave pairing instability is dominant in a wide range of parameter space.
We then argued that this phase will extend to the two other types of mediators.
We have also argued that the previous mean-field results hold for spin-less systems in the weak coupling limit where fluctuations dominate the physics.
Finally, we have identified nearby $s$-wave and $d$-wave singlet pairing phases and identified the critical energy scales for the transition from the Fermi-liquid phase to the pairing phases.
This work demonstrates that systems that do not rely on the exchange repulsion to suppress $s$-wave pairing are capable of triplet superfluidity, and thus opens the possibility of exploring a range of triplet-pairing physics observed in similar systems such as $^3$He and $Sr_2RuO_4$.

{\bf Acknowledgments.} 
This research was supported in part by the NSF under grant DMR-1411345 and by the UC-Lab Fee Research Program under grant LGF-17-476883.

\bibliography{paper}
\end{document}